\documentclass[prl,aps,twocolumn,preprintnumbers,amsmath,amssymb]{revtex4-2}

\usepackage[dvipdfmx]{graphicx}
\usepackage{hyperref}
\usepackage{bm}
\usepackage{color}
\usepackage{graphicx}

\DeclareMathOperator*{\Tr}{{\rm Tr}}

\begin{document}

\title{Interior zeros of supersymmetric indices}

\author{Yu Nakayama}
\email{yu.nakayama@yukawa.kyoto-u.ac.jp}
\affiliation{Center for Gravitational Physics and Quantum Information, Yukawa Institute for Theoretical Physics, Kyoto University, Kitashirakawa Oiwakecho, Sakyo-ku, Kyoto 606-8502, Japan}

\author{Tadashi Okazaki}
\email{tokazaki@seu.edu.cn}
\affiliation{School of Physics and Shing-Tung Yau Center, Southeast University, Yifu Architecture Building, No.2 Sipailou, Xuanwu district, Nanjing, Jiangsu, 210096, China}

\preprint{YITP-26-98}
\date{\today}

\begin{abstract}
A supersymmetric index has interior zeros ($|q|<1$) 
if and only if the arithmetic coefficients $\delta(\nu)$ underlying the supersymmetric zeta function grow exponentially at a rate set by the nearest zero. 
Each $\delta(\nu)$ follows from finitely many $q$-series coefficients, 
so interior zeros are detectable from the expansion alone and obstruct a free or s-confining infrared description, 
as we show for 4d $\mathcal{N}=1$ $SU(2)$ SQCD. 
With a giant graviton expansion interior zeros are a finite $N$ effect, located by the energy of one giant graviton, verified for the $\mathcal{N}=4$ $U(N)$ Schur index.
\end{abstract}

\maketitle

\section{Introduction}
Phase transitions are encoded in the complex zeros of the partition function.
Lee and Yang found these zeros in the complex fugacity \cite{Yang:1952be,Lee:1952ig}, and Fisher in the complex temperature \cite{Fisher:1965zz}.
In each case a phase transition occurs where the zeros approach the real axis in the thermodynamic limit.
For a finite system the partition function is a polynomial, so zeros in the complex plane are unavoidable. 
Its positive coefficients keep them off the positive real axis, and an individual zero is uninformative. 
Only their accumulation onto the real axis in the thermodynamic limit is physical \footnote{For $\mathcal{N}=4$ super Yang-Mills the zeros of the ordinary thermal partition function approach the real temperature axis at the deconfinement transition as $N$ grows \cite{Kristensson:2020nly}.}. 
The supersymmetric index is different in both respects. 
It counts bosonic and fermionic states with opposite signs, so its coefficients are not positive 
and it is a convergent power series, holomorphic in the unit disk rather than a polynomial. 
Such a function need not vanish anywhere, yet the loss of positivity makes interior zeros possible. 
Whether the index has any zeros and where they lie are well-defined questions about the theory at finite rank with no thermodynamic limit required. 
This Letter answers both. 
We give an exact criterion for the existence of interior zeros, a formula for the distance to the nearest one, 
and a count of the interior zeros within any radius, all computable from the $q$-series expansion alone with no closed form required.

The supersymmetric index \cite{Kinney:2005ej,Romelsberger:2005eg} is a
protected quantity that counts BPS operators weighted by their quantum
numbers,
\begin{align}
\label{DEF_ind}
\mathcal{I}(q)&={\Tr}(-1)^F q^{\Delta}
=\sum_n d(n)q^{n/a},
\end{align}
where $F$ is the fermion number, $\Delta$ denotes a protected combination of
conserved charges commuting with chosen supercharges, and $a\in\mathbb{Z}_{>0}$
encodes a possible fractional grading of the spectrum.
It remains invariant under continuous deformations of the theory and therefore
provides a robust probe of the spectrum at strong coupling.
The BPS degeneracy $d(n)$ typically grows at most subexponentially,
\begin{align}
\label{subexp}
\log |d(n)|&=\mathcal{O}(n^{\alpha}), \qquad \alpha<1,
\end{align}
which includes the familiar subexponential growth $\log d(n)\sim n^{(D-1)/D}$ for a $D$-dimensional quantum field theory, 
even without substantial boson-fermion cancellations, as well as oscillatory coefficients. 

In \cite{Nakayama:2025hzr} the \textit{supersymmetric zeta function} was introduced as a BPS spectral function 
extracting the arithmetic information of the spectrum. 
Under (\ref{subexp}) and $\mathcal{I}\neq 0$ on $(0,1)$ it is defined by the Mellin transform
\begin{align}
\label{zeta_Mellin}
\mathfrak{Z}(s,z)
&=\frac{1}{\Gamma(s)}\int_1^{\infty} dq\ \mathrm{PL}[\mathcal{I}(q^{-1})]
q^{-z-1}(\log q)^{s-1},
\end{align}
where
\begin{align}
\label{sind}
\mathrm{PL}[\mathcal{I}(q)]
&=\sum_{d\ge1}\frac{\mu(d)}{d}\log \mathcal{I}(q^d)
=\sum_{\nu\in \frac{1}{a}\mathbb{Z}_{\ge1}}\delta(\nu)q^{\nu}
\end{align}
is the plethystic logarithm \cite{MR1601666} of the index and $\mu(d)$ is the M\"obius function. 
We call $\delta(\nu)$ the \textit{arithmetic coefficients}. 
One convenient representation is the Dirichlet series
\begin{align}
\label{zeta_Dirichlet}
\mathfrak{Z}(s,z)&=\sum_{\nu}\frac{\delta(\nu)}{(\nu+z)^s},
\end{align}
which converges absolutely for sufficiently large $\mathrm{Re}\,s$ provided the arithmetic coefficients $\delta(\nu)$ are polynomially bounded.
An important caveat left unaddressed in \cite{Nakayama:2025hzr} is the regime
in which $\delta(\nu)$ fails to exhibit such controlled growth, invalidating the representation (\ref{zeta_Dirichlet}).

We show that the exponential growth of $\delta(\nu)$ occurs precisely when the supersymmetric index possesses zeros in the open unit disk $|q|<1$. 
The growth rate then equals the inverse modulus of the nearest zero.
Conversely, for free and s-confining theories the plethystic logarithm reduces to a finite sum of single-particle indices, 
so that $\delta(\nu)$ is polynomially bounded and the Dirichlet series is well defined 
\footnote{The absence of interior zeros guarantees only subexponential growth, $\limsup_\nu|\delta(\nu)|^{1/\nu}\le1$, which still permits intermediate growth such as $\delta(\nu)\sim e^{c\sqrt{\nu}}$, for which (\ref{zeta_Dirichlet}) diverges for all $s$. The polynomial bound follows in the free and s-confining case from the finiteness of the plethystic logarithm.}.
The interior zeros of the index thus play the role that Lee-Yang zeros play for the partition function, with the arithmetic coefficients $\delta(\nu)$, 
computable from a $q$-series expansion alone, serving as their quantitative probe.

The zero distribution of the supersymmetric index acquires both physical and mathematical significance.
On the physical side, interior zeros obstruct an infrared description in terms of elementary gauge invariant matter multiplets, such as those
appearing in s-confining dual descriptions of supersymmetric gauge theories \cite{Seiberg:1994pq,Csaki:1996sm,Csaki:1996zb,Aharony:1997gp,Aharony:2013dha}:
such descriptions are naturally expressed as finite plethystic exponentials of single-particle indices and hence admit infinite product representations \cite{Dolan:2008qi,Spiridonov:2008zr}, which a single interior zero rules out.
Mathematically, interior zeros obstruct an infinite product representation altogether, suggesting a new perspective on the classification of special
functions through their infinite product structures, 
with potential applications to the Rogers-Ramanujan type identities \cite{MR1117903} and the Askey-Wilson and Macdonald type integral identities \cite{MR783216,MR1354144}.

\section{Detection of interior zeros}
Let us introduce a uniform grading variable $Q = q^{1/a}$ so that the supersymmetric index admits an ordinary power series expansion
\begin{align}
\label{log_Taylor}
\log \mathcal{I}(q)
= \log \mathcal{I}(Q^a)
= \sum_{m\ge1} L_m Q^m,
\end{align}
where $L_m$ are the Taylor coefficients with respect to the fractional grading variable $Q$ \footnote{We assume a rationally graded spectrum, $\nu\in\frac{1}{a}\mathbb{Z}_{>0}$, as in all examples below. Irrational R-charges, generic after $a$-maximization, require a more careful treatment.}. 
Equivalently, $m \in \mathbb{Z}_{\ge1}$ labels states with charge $\nu = m/a$ in the original $q$-variable.

In this normalization the plethystic logarithm coefficients are defined by M\"{o}bius inversion as
\begin{align}
\label{Mobius_inv}
\delta(\nu)
= \sum_{d\mid \nu a}
\frac{\mu(d)}{d}\,
L_{\frac{\nu a}{d}},
\end{align}
where the sum runs over integers $d$ such that $\nu a / d \in \mathbb{Z}_{\ge1}$. 
Equivalently, one may regard $\delta(\nu)$ as a M\"{o}bius transform of the sequence $\{L_m\}$ under the rescaled lattice $\nu \in \frac{1}{a}\mathbb{Z}_{>0}$.

It follows from the subexponential growth (\ref{subexp}) that
\begin{align}
\limsup_{n\rightarrow \infty}|d(n)|^{1/n}\le 1 ,
\end{align}
and the Cauchy-Hadamard theorem applied to the expansion
$\mathcal{I}(Q^a)=\sum_n d(n)Q^n$ implies that this power series has
radius of convergence at least one. 
In particular, $\mathcal{I}$ is holomorphic in the open unit disk $|Q|<1$ and, 
being given there by a convergent power series, has no poles in this domain. 
Consequently, the only possible singularities of $\log \mathcal{I}$ in $|Q|<1$ arise from zeros of the index 
and the radius of convergence of (\ref{log_Taylor}) equals the distance to the nearest such zero if one exists, and is at least one otherwise.

The asymptotic growth of $L_m$ is thus governed by the zero of $\mathcal{I}$ closest to the origin and, 
as we show in the End Matter, this exponential growth survives the cancellations in the finite M\"obius sum \eqref{Mobius_inv}. 
This leads to the following classification:

\begin{itemize}
   \item \textbf{Non-vanishing indices:} The supersymmetric index has no zeros in the open unit disk. 
   This class includes eta-quotients and more general infinite products. 
   A simple realization is provided by free and s-confining gauge theories. 
   Their infrared physics is described by gauge invariant matter multiplets, so the index factorizes into a finite product of matter multiplet indices, 
   each admitting an infinite product representation. 
   While such theories fall into this zero-free class, the converse is not universally true.
    
    \item \textbf{Indices with zeros in $|q|<1$:} The supersymmetric index develops zeros inside the unit disk. 
    For this class the Taylor coefficients $L_m$ of the logarithm of the index exhibit exponential growth. 
    It excludes s-confining theories as well as free theories, since their indices reduce to finite products of matter multiplet indices.
       
    \item \textbf{Vanishing indices:} The supersymmetric index vanishes everywhere. 
    As the supersymmetric index is a refinement of the Witten index \cite{Witten:1982df}, 
    spontaneous supersymmetry breaking would require the index to vanish, although the converse is not generally true. 
\end{itemize}

An important practical advantage of the arithmetic coefficients $\delta(\nu)$ is that 
the M\"obius inversion (\ref{Mobius_inv}) determines each of them from finitely many Taylor coefficients of $\log\mathcal{I}(q)$. 
No closed-form expression is required: even when the index is given only implicitly, 
e.g. as a multi-dimensional matrix integral, a sufficiently high order $q$-series expansion suffices. 

\section{Zero distance and zero cardinality}
We now introduce two quantitative measures of the interior zeros.
Suppose that the supersymmetric index possesses zeros in the open unit disk and let
\begin{align}
\label{zero_distance}
\rho:=\min\left\{|q|:\mathcal{I}(q)=0,\ |q|<1\right\} \in (0,1)
\end{align}
be the distance from the origin to the nearest such zero, which we call the \textit{zero distance}.

Since $\mathcal{I}(0)=1$, the logarithm is holomorphic in the maximal disk around the origin on which $\mathcal{I}\neq0$, 
so the radius of convergence of (\ref{log_Taylor}) is precisely $\rho_Q:=\rho^{1/a}$, 
the zero distance measured in the grading variable $Q$.
The change of variables maps the open unit disk to itself, so the existence of interior zeros is convention independent, 
while the locations are rescaled as $q_\ast=Q_\ast^{a}$.
Recall that the Cauchy-Hadamard theorem identifies the radius of convergence
$R$ of a power series $\sum_{m\ge0}a_m Q^m$ with the asymptotic growth of its
coefficients through $R^{-1}=\limsup_{m\to\infty}|a_m|^{1/m}$.
Applied to (\ref{log_Taylor}), this yields
\begin{align}
\limsup_{m\to\infty}|L_m|^{\frac{1}{m}}=\rho_Q^{-1}=\rho^{-1/a}>1,
\end{align}
so the Taylor coefficients grow exponentially whenever the index develops an
interior zero, the more rapidly the smaller $\rho$ is.

Crucially, for $\rho<1$ this exponential growth survives the cancellations in the M\"obius sum (\ref{Mobius_inv}). 
The number of divisors grows at most sub-polynomially, 
and every term with $d\ge2$ involves $L_{m/d}$ with $m/d\le m/2$, 
so all of them together are exponentially suppressed against the $d=1$ term (End Matter).
Taking into account the spectral grading $\nu=m/a$, we obtain the exact equivalence of the growth rates:
\begin{align}
\label{growth_zerodistance}
\limsup_{\nu\to\infty}|\delta(\nu)|^{\frac{1}{\nu}}
= \left( \limsup_{m\to\infty}|L_m|^{\frac{1}{m}} \right)^a = \rho^{-1} > 1,
\end{align}
or equivalently $\limsup_{m\to\infty}|\delta(m/a)|^{1/m}=\rho_Q^{-1}=\rho^{-1/a}$ 
in terms of the integer expansion order $m$ in $Q$ used in all numerical analyses below.
The zero distance therefore provides a quantitative measure of the obstruction
to the convergence of the Dirichlet series (\ref{zeta_Dirichlet}).

The interior zeros can also be counted.
Let
\begin{align}
\label{zero_cardinality}
\mathcal{N}(r):=\#\left\{q:\mathcal{I}(q)=0,\ |q|\le r\right\},\qquad r<1,
\end{align}
counted with multiplicity, and call $\mathcal{N}(r)$ the \textit{zero
cardinality}.
The argument principle expresses it as the winding number
$\frac{1}{2\pi i}\oint_{|q|=r}d\ln\mathcal{I}$, which is again computable from
a high-order truncation of the series.
Jensen's formula turns the count into a sum rule. 
Since $\mathcal{I}(0)=1$,
\begin{align}
\label{Jensen}
\frac{1}{2\pi}\int_0^{2\pi}\!d\theta\,
\log\big|\mathcal{I}(re^{i\theta})\big|
=\int_0^{r}\frac{dt}{t}\,\mathcal{N}(t).
\end{align}
A zero-free index therefore has a vanishing circle average of $\log|\mathcal{I}|$ at every radius, 
no matter how fast the index grows in particular directions.
A non-zero average signals interior zeros, and its growth as $r\to1$ counts them.

\section{4d $\mathcal{N}=1$ $SU(2)$ gauge theories}
\begin{figure}[t]
\centering
\includegraphics[width=\columnwidth]{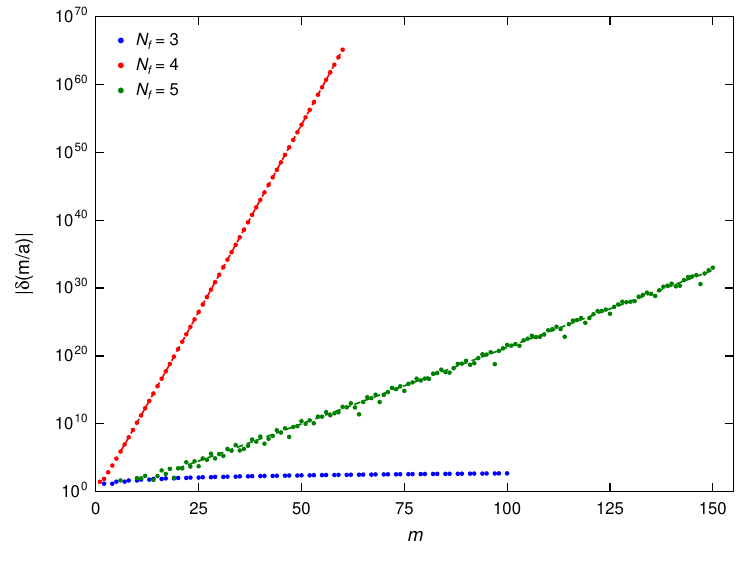}
\caption{Arithmetic coefficients $|\delta(m/a)|$ (log scale) versus expansion
order $m$ for 4d $\mathcal{N}=1$ $SU(2)$ gauge theories with $N_f=3$ (blue), $4$ (red), and $5$ (green). 
For the s-confining $N_f=3$ theory the coefficients follow the free meson content and grow only linearly, 
whereas for $N_f=4,5$ they grow exponentially; the dashed lines show the fitted growth $|\delta|\propto\rho_Q^{-m}/m$ for $N_f=4$ and $5$.}
\label{fig_4dN1su2}
\end{figure}

Our first example is the 4d $\mathcal{N}=1$ $SU(2)$ gauge theory with $2N_f$ fundamental chiral multiplets.
The superconformal index \cite{Kinney:2005ej,Romelsberger:2005eg} depends on two fugacities $p$ and $q$, and we set $p=q$. 
The chiral multiplets have superconformal R-charge $1-\frac{2}{N_f}$ \cite{Seiberg:1994pq,Intriligator:2003jj}. 
The spectrum is graded by $a=N_f$ for $N_f=3,5$ and by $a=1$ for $N_f=4$. 
No closed form is known for general $N_f$. 
The index is given by a matrix integral, which makes high-order coefficients non-trivial to obtain. 
Using exact integer arithmetic we expand it in $Q$ and compute the arithmetic coefficients $\delta(\nu)$ from (\ref{Mobius_inv}) (End Matter). 
The results are summarized in Fig.~\ref{fig_4dN1su2}.
The asymptotic behavior of $|\delta(\nu)|$ diagnoses the infrared theory. 
A bounded or subexponentially growing sequence is necessary for a free or s-confining description, which guarantees a factorizable infinite product, 
while exponential growth rigorously obstructs any such dual. 
For $N_f=3$ the theory s-confines into fifteen mesons of R-charge $\frac{2}{3}$ \cite{Seiberg:1994pq,Intriligator:1995ne}, 
and the matrix integral reduces to the free-meson index \cite{MR1846786,Dolan:2008qi,Spiridonov:2008zr}. 
Our expansion reproduces this exactly. 
The coefficients match the single-particle content of the fifteen mesons, 
$\delta(\tfrac{2}{3}+k)=15(k+1)$ and $\delta(\tfrac{4}{3}+k)=-15(k+1)$, at every order $m\le100$, growing only linearly, so the growth rate is one, 
in accord with the manifestly zero-free product form of the index.

At $N_f=4$ and $5$ the coefficients grow exponentially from the lowest orders.
Both indices therefore vanish inside the disk, so neither infrared can be free or s-confining, 
consistent with the interacting fixed points of the conformal window. 
The location of the nearest zero controls how fast the growth rate estimate converges. 
For $N_f=4$ it is real, negative, and well isolated at $Q_\ast=-0.0767$, 
whereas for $N_f=5$ the nearest zeros are a complex conjugate pair at $|Q_\ast|=0.5858$ with a second pair only about $1\%$ further out, 
so $|\delta(\nu)|$ oscillates and the rate converges more slowly.
Either way the series alone locates the zeros, separating these interacting fixed points from the zero-free s-confining theory.

\section{Zeros of giant gravitons}
\label{sec_GG}
When the index of a theory with a holographic dual admits a giant graviton expansion \cite{Arai:2020qaj,Gaiotto:2021xce}, 
its interior zeros are a finite $N$ effect produced by the wrapped branes, 
and the expansion also fixes where the nearest zero lies. 
It writes the finite $N$ index as its large $N$ limit times a series of corrections, 
the $n$-th suppressed by $Q^{nE}$ because it describes $n$ branes of charge $E$, 
the energy of a BPS brane set by the rank $N$ up to the normalization of the fugacity. 
The leading correction, from the $n=1$ sectors, is therefore of the form $-CQ^{E}$, 
the negative sign reflecting that these corrections remove states present at large $N$. 

When the internal space provides two cycles for the giant gravitons to wrap 
and the flavored index carries enough independent fugacities to grade the two wrapping numbers separately, 
the index admits a double sum giant graviton expansion 
(e.g. for the suitable specializations of the indices of the D3-branes \cite{Arai:2020qaj} and the M2- and M5-branes \cite{Hayashi:2024aaf}) 
and the coefficient $C$ acquires a simple origin. 
The branes wrapping the two cycles then carry opposite flavor weights, contributing $x^{N}$ and $y^{N}$ times giant graviton indices that, 
as often happens, are separately singular as the flavor fugacities degenerate, $x,y\to Q$ \cite{Imamura:2022aua}. 
The poles cancel in the sum, leaving a difference quotient $(x^{N}-y^{N})/(x-y)$ which becomes $Nx^{N-1}$: 
the limit brings the exponent down as the coefficient and identifies $C=N$ with an energy \footnote{With more cycles the sectors of unit wrapping number are more numerous (see e.g. \cite{Imamura:2021ytr,Arai:2020uwd}), and their poles should cancel into a higher divided difference. We do not pursue this here.}.

An interior zero first becomes possible where this correction cancels the vacuum, $C|Q|^{E}\sim1$.
At large $N$ the higher wrapping sectors, suppressed by further powers of $Q^{E}$, 
do not restore the cancellation, and $C$ differs from $E$ only by a shift of relative order $1/E$, so that
\begin{align}
\label{giant_law}
-E\log\rho_Q=\log E+\mathcal{O}(1) .
\end{align}
The nearest zeros then sit at $\rho_Q\simeq E^{-1/E}$, that is $1-\rho_Q\simeq(\log E)/E$, approaching the unit circle as $E$ grows.
Conversely, since the growth of $|\delta(\nu)|$ determines
$\rho_Q$ through (\ref{growth_zerodistance}), the $q$-series expansion alone determines the energy of the giant graviton to leading order, 
which we now test against the $\mathcal{N}=4$ $U(N)$ Schur index. 

\section{4d $\mathcal{N}=4$ super Yang-Mills theories}
\label{sec_N4}
Our second example is the Schur index \cite{Gadde:2011ik,Gadde:2011uv} of 4d $\mathcal{N}=4$ super Yang-Mills theory with gauge group $U(N)$, 
for which a closed form is known \cite{Bourdier:2015wda} (also see \cite{Beem:2021zvt,Pan:2021mrw,Huang:2022bry,Hatsuda:2022xdv}), 
\begin{align}
\label{Schur_uN}
\mathcal{I}^{\textrm{$\mathcal{N}=4$ $U(N)$}}&=\vartheta_4(0,Q)^{-1}S_N(Q), \\
\label{SN_series}
S_N(Q)&=\sum_{n\ge0}(-1)^n
\left[\tbinom{N+n}{N}+\tbinom{N+n-1}{N}\right]Q^{\,nN+n^2},
\end{align}
with $Q=q^{1/2}$ and $\vartheta_4(0,Q)^{-1}=\prod_{n\ge1}\big[(1-Q^{n})(1-Q^{2n-1})\big]^{-1}$.
The prefactor is the large $N$ index, which counts the graviton Fock space 
and has neither zeros nor poles in $|Q|<1$, so every interior zero of the index is a zero of $S_N(Q)$.
The series $S_N$ can be viewed as the unflavored limit of the giant graviton expansion \cite{Arai:2020qaj,Gaiotto:2021xce} (also see \cite{Murthy:2022ien,Beccaria:2024szi}). 
Its $n$-th term collects the sectors of total wrapping number $n$, the power $Q^{nN}$ being the classical weight of $n$ branes \cite{Arai:2020qaj}. 
The leading correction $-(N+2)Q^{N+1}$ realizes the general structure of Sec.~\ref{sec_GG}, 
with both the exponent and the coefficient equal to the weight $N$ of a single brane up to a shift. 
The arithmetic coefficients and the zeros are computable to high precision, which makes this family a stringent test of (\ref{giant_law}).

\begin{figure}[t]
\centering
\includegraphics[width=\columnwidth]{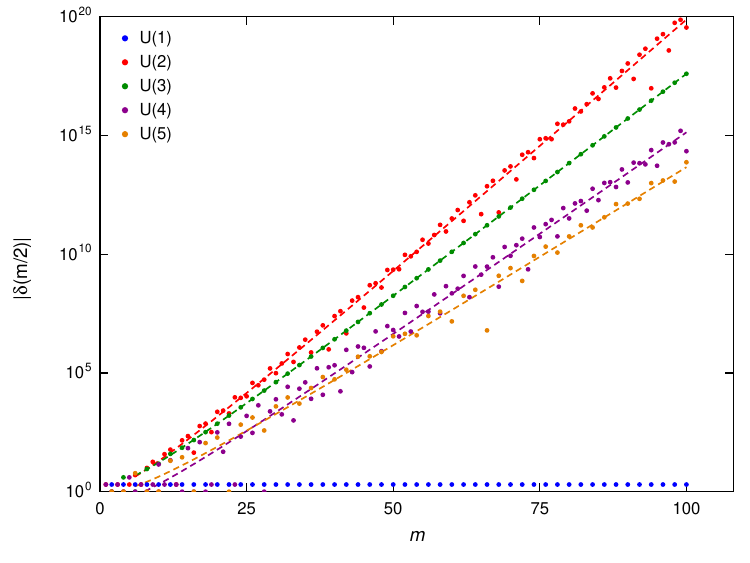}
\caption{Arithmetic coefficients $|\delta(m/2)|$ (log scale) for the Schur indices of 4d $\mathcal{N}=4$ $U(N)$ theories up to $m=100$, with the fits described in the End Matter (dashed). For odd $N$ only even orders are shown, since the zero-carrying factor $S_N(Q)$ is even in $Q$. The flat $U(1)$ trajectory reflects the free factorized index. The exponential growth for all $N\ge2$ signals interior zeros in $|q|<1$, and the scatter about each fit reflects the phase of the nearest zero.}
\label{fig_4dN4}
\end{figure}

Figure \ref{fig_4dN4} shows $|\delta(m/2)|$ up to $m=100$ for $N\le5$.
The free $U(1)$ theory has bounded coefficients, $\delta(m/2)=2(-1)^{m+1}$,
matching the zero-free product form of its index.
Every non-Abelian theory grows exponentially instead, reaching
$|\delta(m/2)|\simeq1.2\times10^{216}$ at $m=1000$ for $U(2)$, so each one has interior zeros. 

We estimate the zero distance from this growth,
$\rho_{\delta}:=\big(\limsup_m|\delta(m/2)|^{1/m}\big)^{-2}$, which
(\ref{growth_zerodistance}) equates with $\rho$.
Table \ref{tab_N4} confirms this against the zeros computed directly from $S_N$. 
\begin{table}[t]
\caption{Nearest zeros of the 4d $\mathcal{N}=4$ $U(N)$ Schur indices in the $Q=q^{1/2}$ plane. The estimate $\rho_{\delta}^{1/2}$ is the inverse of the growth rate measured from $\delta(m/2)$ with $m\le1000$. This fit is accurate to about $0.01\%$.}
\label{tab_N4}
\begin{ruledtabular}
\begin{tabular}{lcccc}
 & $\limsup_m|\delta(m/2)|^{\frac1m}$ & $\rho_{\delta}^{1/2}$ & $|Q_\ast|$ & $\arg Q_\ast$ \\
\hline
$U(2)$ & $1.6558$ & $0.60393$ & $0.603936$ & $\pm117.77^\circ$ \\
$U(3)$ & $1.5594$ & $0.64129$ & $0.641287$ & $\pm90^\circ$ \\
$U(4)$ & $1.4816$ & $0.67494$ & $0.674992$ & $\pm72.87^\circ$ \\
$U(5)$ & $1.4217$ & $0.70336$ & $0.703328$ & $\pm61.36^\circ,\ \pm118.64^\circ$ \\
\end{tabular}
\end{ruledtabular}
\end{table}
The nearest zeros are simple, 
none on the positive real axis as we can prove from the matrix integral expression: a single conjugate pair, except for $U(5)$, 
where the evenness of $S_N$ adds its image under $Q_\ast\to-Q_\ast$ at $\pm61.36^\circ$. 
The zero cardinality accordingly vanishes for $r<\rho$, jumps by two there ($U(5)$: four), and grows without bound as $r\to1$.

A single cancellation accounts for these zeros, as anticipated by (\ref{giant_law}).
Truncating (\ref{SN_series}) after the $n=1$ term gives $S_N=1-(N+2)\,Q^{N+1}+\mathcal{O}(Q^{2N+4})$, so the index can first vanish
where the unit wrapping sectors cancel the vacuum contribution. 
The truncated condition $(N+2)\,Q^{N+1}=1$ has $N+1$ solutions, equally spaced on the circle of radius $(N+2)^{-1/(N+1)}$, 
and the zeros of the full series stay close to the complex ones, so all the angles in Table \ref{tab_N4} lie within $3^\circ$ of $2\pi j/(N+1)$ for an integer $j$.

Because (\ref{giant_law}) is an asymptotic law, we have tested it by localizing the nearest zeros directly up to $N=10^4$, 
where the combination predicted to be constant indeed settles (End Matter). 
The distance of the zeros from the unit circle therefore shrinks as $(\log N)/N$, 
set by the energy of a single giant graviton, confirming the general law of Sec.~\ref{sec_GG} in a controlled example.

At $N=\infty$ the sum (\ref{SN_series}) collapses to its first term and the index reduces to the zero-free prefactor $\vartheta_4(0,Q)^{-1}$, 
the index of the graviton Fock space of the classical supergravity background. 
The interior of the disk is then free of zeros.

\section{Conclusions}
We have related the zeros of a supersymmetric index to the growth of the \textit{arithmetic coefficients} $\delta(\nu)$.
Zeros inside the unit disk exist precisely when the coefficients $\delta(\nu)$ grow exponentially, and the growth rate measures the modulus of the nearest zero.
This relation turns a question about the zeros of the index into a computation with finitely many series coefficients. 
As an infrared diagnostic it rules out free and s-confining descriptions for $SU(2)$ SQCD with $N_f=4,5$, 
where we also locate the nearest zeros, whereas at $N_f=3$ the coefficients reproduce the single-particle content of the confined mesons. 
For the $\mathcal{N}=4$ Schur indices the same growth rate locates the interior zeros, 
which lie where a single giant graviton contribution cancels the vacuum contribution and whose distance obeys the large $N$ law (\ref{giant_law}). 

Our results open up several promising directions.
Since the arithmetic coefficients $\delta(\nu)$ require only a low order $q$-series expansion rather than a closed-form expression, 
our diagnostic applies directly to broader classes of supersymmetric indices, including indices of 3d and 6d SCFTs and protected sectors with defect operators. 
It would be interesting to follow the zero distance across families of theories and to connect the finite $N$ zeros with the large $N$ phase transitions of the complexified index \cite{Copetti:2020dil,Goldstein:2020yvj,Deddo:2025jrg}. 
Mathematically, the interplay between the interior zeros of matrix integrals and the asymptotics of their arithmetic coefficients warrants rigorous formulation, 
and may offer a new classification of special functions, such as those appearing in Rogers-Ramanujan type identities \cite{MR1117903} 
and Macdonald type integral identities \cite{MR783216,MR1354144}, through their infinite product structures. 
Finally, the divergence of the Dirichlet series caused by interior zeros invalidates the series description in terms of the arithmetic coefficients $\delta(\nu)$, 
forcing us to rely on more general Mellin representation of the supersymmetric zeta function \cite{Nakayama:2025hzr} for further investigations. 

\begin{acknowledgments}
YN is supported in part by JSPS KAKENHI Grant Number 21K03581 and 26K00699. 
TO was supported by the Startup Funding no.\ 4007012317 of Southeast University. 
\end{acknowledgments}

\bibliographystyle{utphys}
\bibliography{ref}

\appendix
\section{Proof of the growth relation}
\label{app_proof}
We use the notation of the main text and set $G:=\limsup_{m\to\infty}|L_m|^{1/m}$.

The first step determines the radius of convergence of (\ref{log_Taylor}). 
The bound (\ref{subexp}) gives $\limsup_n|d(n)|^{1/n}\le1$, so by the Cauchy-Hadamard theorem $\mathcal{I}$ is holomorphic in $|Q|<1$.
Let $D$ be the largest open disk centered at the origin containing no zero of $\mathcal{I}$. 
Since $D$ is simply connected and $\mathcal{I}(0)=1$, the principal branch of $\log\mathcal{I}$ is holomorphic on $D$ and (\ref{log_Taylor}) converges there.
If $\mathcal{I}$ has interior zeros, the boundary of $D$ passes through a nearest zero at $|Q|=\rho_Q$. 
Near such a point $\log|\mathcal{I}|$ diverges, while the sum of a power series is bounded on compact subsets of its disk of convergence.
The radius of convergence therefore equals $\rho_Q$ exactly, and is at least one when there are no interior zeros.
Cauchy-Hadamard then yields $G=\rho_Q^{-1}>1$ in the first case and $G\le1$ in the second.

The second step shows that the M\"obius sum preserves the growth. 
If $G>1$ then $\limsup_m|\delta(m/a)|^{1/m}=G$, and if $G\le1$ then $\limsup_m|\delta(m/a)|^{1/m}\le1$.
The number of divisors of $m$ is at most $2\sqrt{m}$, since they pair up as $(d,m/d)$ with $\min(d,m/d)\le\sqrt{m}$.
Fix $\varepsilon>0$, set $M=\max(G+\varepsilon,1)$, and let $C_\varepsilon$ be such that $|L_k|\le C_\varepsilon(G+\varepsilon)^k$ for all $k\ge1$.
Every term of (\ref{Mobius_inv}) is bounded by $C_\varepsilon M^m$, 
because $|\mu(d)/d|\le1$ and $(G+\varepsilon)^{m/d}\le M^{m}$, 
so $|\delta(m/a)|\le2\sqrt{m}\,C_\varepsilon M^m$ and $\limsup_m|\delta(m/a)|^{1/m}\le M$ for every $\varepsilon$. 
This gives the upper bounds in both statements. 
For the lower bound let $G>1$ and choose $\varepsilon$ with $(G+\varepsilon)^{1/2}<G-\varepsilon$, possible because $\sqrt{G}<G$.
Along a subsequence with $|L_{m_k}|\ge(G-\varepsilon)^{m_k}$, 
the terms with $d\ge2$ have $m_k/d\le m_k/2$ and contribute at most $2\sqrt{m_k}\,C_\varepsilon(G+\varepsilon)^{m_k/2}$. 
Thus we have
\begin{align}
|\delta(m_k/a)|&\ge(G-\varepsilon)^{m_k}-2\sqrt{m_k}\,C_\varepsilon(G+\varepsilon)^{m_k/2}
\nonumber\\
&\ge\tfrac12(G-\varepsilon)^{m_k}
\end{align}
for large $k$, so $\limsup_m|\delta(m/a)|^{1/m}\ge G-\varepsilon$ and the equality follows.

With $\nu=m/a$ we have $|\delta(\nu)|^{1/\nu}=(|\delta(m/a)|^{1/m})^a$, so interior zeros give $\limsup_\nu|\delta(\nu)|^{1/\nu}=G^a=\rho^{-1}>1$, 
and their absence gives $\limsup_\nu|\delta(\nu)|^{1/\nu}\le1$.
This is (\ref{growth_zerodistance}).

\section{Numerical methods}
\label{app_num}
Reading $\rho$ from the growth of $\delta(\nu)$ calls for care.
A simple nearest zero makes $\log\mathcal{I}$ logarithmically singular, 
so $|\delta(\nu)|\sim C\rho^{-\nu}/\nu$, and the bare sequence $|\delta(\nu)|^{1/\nu}$ approaches $\rho^{-1}$ only as $1-(\log\nu)/\nu$. 
Removing the $\nu$ restores the exponential rate, and a linear fit of $\log(\nu|\delta(\nu)|)$ over the upper half of the range determines $\rho$.
Its accuracy depends on how well the nearest zero is isolated from the others.

The $SU(2)$ superconformal indices are known only as matrix integrals.
We expand the integrand in $Q$ with exact integer coefficients and project onto gauge invariants with the $SU(2)$ Haar measure, 
order by order, to $m_{\max}=100$, $60$ and $150$ for $N_f=3,4$ and $5$. 
The resulting $\delta(\nu)$ come out integers, as degeneracies must. 
The fit reaches $0.001\%$ for $N_f=4$ at $m=60$, because that zero is real and well separated, 
whereas the $N_f=5$ zeros are complex and nearly degenerate and the fit converges more slowly. 
Since only a truncated series is available, 
the zeros are located by taking all roots of the truncation and keeping those that are stable against changing the truncation order.

For the $\mathcal{N}=4$ Schur indices the closed form (\ref{Schur_uN}) gives the expansion to $m=1000$, and the fit returns $\rho$ to about $0.01\%$. 
The bare sequence would still overshoot by $0.7\%$ at $m=1000$ for $U(2)$.
For odd $N$ the factor $S_N(Q)$ carrying the zeros is even in $Q$, so the growth appears at even orders only and the fit is restricted to them. 
Here the zeros are obtained from the logarithmic derivative $\partial_Q\log S_N$, whose diagonal Pad\'e approximants have poles at the zeros of $S_N$. 
Newton's method refines each pole to a zero, 
and the argument principle confirms it through a unit winding of a small contour and returns $\mathcal{N}$ as the winding along a circle of fixed radius.
At large $N$ the zeros are reached by Newton's method based at the angles $2\pi j/(N+1)$, 
with $S_N$ and its derivative evaluated through their logarithms to control the large magnitudes.
Table \ref{tab_ring} extends the test of (\ref{giant_law}) to $N=10^4$.
The combination $|w_N|=N\rho_Q^{\,N+1}$ measures the law, 
since $-\log |w_N|$ is exactly the quantity that (\ref{giant_law}) predicts to approach a constant. 
It does so slowly and not monotonically, rising past its limit before decreasing again, so extrapolation from moderate rank overestimates it. 
\begin{table}[t]
\caption{Nearest zeros of $S_N(Q)$ at large $N$ and the combination
$|w_N|=N\rho_Q^{\,N+1}$. The listed $\rho_Q$ are rounded; 
$|w_N|$ is computed from the full precision values.}
\label{tab_ring}
\begin{ruledtabular}
\begin{tabular}{lccc}
$N$ & $\rho_Q$ & $\arg Q_\ast$ & $|w_N|$ \\
\hline
$10$ & $0.782593$ & $\pm98.01^\circ$ & $0.6744$ \\
$50$ & $0.920621$ & $\pm70.70^\circ$ & $0.7364$ \\
$10^2$ & $0.952585$ & $\pm110.44^\circ$ & $0.7401$ \\
$10^3$ & $0.992823$ & $\pm68.70^\circ$ & $0.7390$ \\
$10^4$ & $0.999049$ & $\pm68.57^\circ$ & $0.7380$ \\
\end{tabular}
\end{ruledtabular}
\end{table}

The numerical results and the associated Mathematica notebook are provided as ancillary files in the arXiv submission.

\section{Giant graviton zeros and the Sokal conjecture}
\label{app_sokal}
For the $\mathcal{N}=4$ $U(N)$ Schur index the constant in (\ref{giant_law}) can be determined, and it is controlled by a single function.
The one-giant sector has weight $w_N=N Q^{N+1}$, including its degeneracy $N$. 
We take $N\to\infty$ holding $w_N$ and the angle $\theta=\arg Q$ fixed, and write $w$ for the limiting value of $w_N$.
Using the identity $nN+n^2=n(N+1)+n(n-1)$ specific to (\ref{SN_series}), in
this limit the coefficient of the $n$-th term approaches $N^n/n!$ while
$Q^{n(n-1)}$ tends to the pure phase $e^{i\theta n(n-1)}$, so the zero condition $S_N=0$ reduces to
\begin{align}
\label{Phi}
\Phi(w,\theta)=\sum_{n\ge0}\frac{(-w)^n}{n!}\,e^{i\theta\,n(n-1)}=0.
\end{align}
The phase factor $e^{i\theta n(n-1)}$ is what produces the zeros. 
At $\theta=0$ it is absent, $\Phi(w,0)=e^{-w}$ has no zeros, and none appear. 
A non-zero $\theta$ switches it on together with the zeros. 
Writing $w_*(\theta)$ for the zero of $\Phi$ nearest the origin, the nearest
zero of the index sits at the angle that brings it closest, so $|w_N|$ converges to
\begin{align}
\label{winf}
w_\infty=\min_\theta|w_*(\theta)|=0.7378\ldots
\end{align}
at $\theta_*=68.56^\circ$ and, 
since $n(n-1)$ is even, equally at $180^\circ-\theta_*=111.44^\circ$, between whose neighborhoods the angles in Table \ref{tab_ring} alternate at large $N$.
This determines the constant in (\ref{giant_law}):
\begin{align}
\label{ringlaw}
-(N+1)\log\rho_Q=\log N-\log w_\infty+\mathcal{O}\!\left(\frac{\log N}{N}\right).
\end{align}
Thus $w_\infty$ is the universal value at which the one-giant graviton sector first cancels the vacuum in the large $N$ limit, 
and it fixes the rate at which the zeros approach the unit circle.
The function $\Phi$ is the deformed exponential $\sum_n x^ny^{\binom{n}{2}}/n!$ at $x=-w$ and $y=e^{2i\theta}$, 
whose zeros are conjectured by Sokal to be simple and of distinct moduli for $0<|y|\le 1$ (see e.g. \cite{Dyachenko:2013,MR3810255}).
Determining $w_\infty$ is the problem of finding the zero of this function nearest the origin on the unit circle $|y|=1$. 
Whether the interior zeros of other indices with giant graviton expansions map onto zero distribution problems in classical analysis, 
as they do here, is an interesting question.

\end{document}